\documentclass{iopart}

\renewcommand\vec[1] {\ensuremath{\mathbf{#1}}}

\newcommand\ppar{\ensuremath{p_\parallel}}
\newcommand\pperp{\ensuremath{p_\perp}}
\newcommand\Tpar{\ensuremath{T_\parallel}}
\newcommand\Tperp{\ensuremath{T_\perp}}

\newcommand\Div[1]{\ensuremath{\nabla\cdot#1}}
\newcommand\Curl[1]{\ensuremath{\nabla\times#1}}
\newcommand\Grad{\ensuremath{\nabla}}
\newcommand\Cross{\ensuremath{\times}}
\newcommand\Abs[1]{\ensuremath{\left |#1\right |}}

\newcommand\Equation[1]{Eq.~(\ref{#1})}
\newcommand\FIGURE[1]{Fig.~\ref{#1}}
\newcommand\Eqns[2]{Eqs.~(#1)~and~(#2)}
\newcommand\EqnsMany[2]{Eqs.~(#1)-(#2)}

\newcommand\Exp[1]{\ensuremath{\exp{\left(#1\right)}}}
\newcommand\Ln[1]{\ensuremath{\log{\left(#1\right)}}}
\newcommand\Sum[3]{\ensuremath{\sum^{#3}_{#1=#2}}}
\newcommand\Norm[1]{\ensuremath{\|#1\|}}
\newcommand\Iter[1]{\ensuremath{^{(#1)}}}
\newcommand\Deby[1]{\ensuremath{\frac{d}{d#1}}}
\newcommand\PsiNorm{\ensuremath{\bar{\psi}}}
\newcommand\Brackets[1]{\ensuremath{\left(#1\right)}}
\newcommand\SBrackets[1]{\ensuremath{\left[#1\right]}}
\newcommand\E[1]{\ensuremath{\times~10^{#1}}}

\newcommand\IE{i.e.:~}
\newcommand\Solovev{Soloviev~}
\newcommand\Alfven{Alfv\'{e}n}
\newcommand\EG{e.g.:~}

\newcommand\text[1]{\ensuremath{\rm{#1}}}
\newcommand\dyad[1] {\ensuremath{\rm{#1}}}
\usepackage{graphicx}

\newcommand\Newline{\ensuremath{\nonumber \\}}

\begin{document}
\title{EFIT tokamak equilibria with toroidal flow and anisotropic pressure using the two-temperature guiding-centre plasma}
\date{\today}
\begin{abstract}
A new force balance model for the EFIT magnetohydrodynamic equilibrium technique for tokamaks is presented which includes the full toroidal flow and anisotropy changes to the Grad-Shafranov equation. The free functions are poloidal flux functions and all non-linear contributions to the toroidal current density are treated iteratively. The parallel heat flow approximation chosen for the model is that parallel temperature is a flux function and that both parallel and perpendicular pressures may be described using parallel and perpendicular temperatures. This choice for the fluid thermodynamics has been shown elsewhere to be the same as a guiding centre kinetic solution of the same problem under the same assumptions.  The model reduces identically to the static and isotropic Grad-Shafranov equation in the appropriate limit as different flux functions are set to zero. An analytical solution based on a modified \Solovev
solution for non-zero toroidal flow and anisotropy is also presented.

The force balance model has been demonstrated in the code EFIT TENSOR, a branch of the existing code EFIT++. Benchmark results for EFIT TENSOR are presented and the more complicated force balance model is found to converge to force balance  similarly to the usual EFIT model and with comparable speed.
\end{abstract}
\author{M. Fitzgerald}
\ead{Michael.Fitzgerald@anu.edu.au}
\address{Research School of Physical Sciences and Engineering, Australian National University, 0200
ACT, Australia}
\author{L.C. Appel}
\address{EURATOM/CCFE Fusion Association, Culham Science Centre, Abingdon, Oxfordshire,
OX14 3DB, UK}
\author{M.J. Hole}
\address{Research School of Physical Sciences and Engineering, Australian National University, 0200
ACT, Australia}

\section{Introduction}
The macroscopic equations of magnetohydrodynamics (MHD) provide the basic starting point for an understanding of plasma physics in a modern tokamak experiment. Good knowledge of the total equilibrium force balance provides information about the magnetic topology and the plasma thermodynamic variables so that more detailed stability and transport treatments can be pursued. The success of the tokamak concept has been possible, in part, due to the simple models which support inference of the plasma configuration from  incomplete measurements of related parameters. Most basic physical variables can only be measured indirectly on a fusion experiment through a sophisticated and expensive set of diagnostics. Good physical models of the configuration are essential, particularly for future power stations which cannot accommodate complex diagnostics programmes. 

Modern tokamak experiments contain a significant portion of fast ions \cite{Fasoli2007} resulting from heating processes such as neutral beam injection (NBI) and ion-cyclotron resonance heating (ICRH) which can rotate the plasma and also produce highly anisotropic fast particle pressures \cite{Cooper1980,Hole2011}. Both of these effects can result in pressure surfaces that are no longer magnetic surfaces and can significantly alter the density profile and magnetic topology (see, for example,  \cite{Iacono1990,Guazzotto2004,Hole2009,Fitzgerald2011,Pustovitov2010}).

The tokamak equilibrium reconstruction code EFIT  \cite{Lao1985} has served as the de-facto standard technique to infer equilibrium from experimental diagnostics and there have been many different code implementations of this technique. EFIT solves the MHD force balance for static and isotropic pressure, although toroidal flow (see for example \cite{Lao2005} and \cite{Appel}) and anisotropy \cite{Zwingmann2001} have been previously included in an approximate fashion through modifications to the pressure term (see Appendix). For the particular phenomenon of toroidal flow, it can be shown that the deviation of density from magnetic surfaces obeys an exponential scaling. Existing EFIT implementations such as \cite{Lao2005} make an asymptotic expansion of this exponential dependance in low toroidal Mach number, thereby removing the non-linear dependance for convenient implementation in linear inversion. The magnitude of the rotation in these models is quantified with a new `rotational-pressure' flux function. In addition to assuming low Mach number, this approximation is only valid to lowest order in anisotropy.

In this paper, we describe an extension of the basic EFIT algorithm to include the fully non-linear toroidal flow and anisotropy contributions to the 2-D plasma equilibrium problem. This is in contrast with the low Mach number  \cite{Lao2005} and small `anisotropic $\beta$'  \cite{Zwingmann2001} EFIT implementations attempted previously. The physical model is based on the guiding-centre plasma (GCP) formalism  \cite{Grad1966}  as derived by Dobrott and Greene \cite{Dobrott1970} for a two-temperature anisotropic plasma model \cite{Iacono1990}.  This new algorithm has been demonstrated in EFIT TENSOR, a code branch created by modifying an existing EFIT implementation used currently for the MAST tokamak (known as `EFIT++'  \cite{Appel}).  We will present analytical and Mega Ampere Spherical Tokamak (MAST) \cite{Cox1999} test case equilibria produced by EFIT TENSOR and demonstrate correct numerical convergence to force balance.
\section{Basic equations and assumptions}
We are concerned with the system of macroscopic equilibrium equations, based on the guiding-centre plasma and the ideal MHD Ohm's law, given in natural units as \cite{Grad1966,Dobrott1970}
\begin{eqnarray}
\rho(\vec{u}\cdot\nabla\vec{u})+\nabla\cdot\dyad{P}= \vec{J} \times \vec{B}  \label{force}\\
\dyad{P}=\pperp\dyad{I}+\Delta\vec{B}\vec{B}, \Delta\equiv\frac{\ppar-\pperp}{B^2} \label{pdyad} \\
\Div{\vec{B}}=0 \label{monopole}\\ 
\Div{\rho\vec{u}}=0 \label{mass}\\ 
\Curl{\vec{B}}=\vec{J} \label{ampere}\\ 
\vec{E}+\vec{u}\Cross\vec{B}=0  \label{superconductive}\\
\Curl{\vec{E}}=0 \label{induction}
\end{eqnarray}
The single fluid pressure dyad $\dyad{P}$ and its components $\ppar,\pperp$, the single fluid velocity $u$ and single fluid mass density
$\rho$ have textbook definitions (\EG\cite{Bellan2006}) in terms of each individual fluid equation for each species present in the plasma, which in turn are moments of the guiding centre plasma equation for each species.
Divergence-less vector fields (\Eqns{\ref{monopole}}{\ref{mass}}) in a 2-d axisymmetric cylindrical system $(R,\phi,Z)$ permit the covariant representations
\begin{eqnarray}
\vec{B}=\Grad{\psi}\Cross\Grad\phi+R B_\phi \Grad\phi \label{covB} \\
\vec{\rho\vec{u}}=\Grad{\psi_M}\Cross\Grad\phi+ R\rho u_\phi \Grad \phi \label{covrhou}
\end{eqnarray}
in terms of poloidal stream functions $\psi$ and $\psi_M$ and toroidal components of field $B_\phi $ and flow $u_\phi$. Combining \Eqns{\ref{covB}}{\ref{covrhou}} with \Eqns{\ref{superconductive}}{\ref{induction}} gives the well-known `frozen-in' condition for magnetic field lines in axisymmetric cylindrical geometry
\begin{equation}
\vec{u}=\frac{\psi_M'(\psi)}{\rho}\vec{B}+R^2\Omega(\psi)\Grad{\phi} \label{flowfrozen}
\end{equation}
which specifies that, to conserve magnetic flux through a given fluid element, all macroscopic ideal flow must occur parallel to the magnetic field or in a symmetry direction. Under these assumptions, flow in a poloidal direction can only manifest as the projection of flow in the parallel direction and an additional toroidal angular velocity $\Omega$ is constant on poloidal flux surfaces. In this study, we neglect all poloidal flows setting $\psi_M'(\psi)=0$.

To close the system of equations, an energy equation is required. Ignoring resistivity or other dissipation, the work done against the pressure equals the change in plasma energy U for a reversible process. Taking $\vec{u} \cdot \Div \dyad{P}$ gives terms in $\Div \vec{u}$ and $\vec{B}\cdot \nabla \vec{u}$ which may be related to the convective derivative $d/dt=\vec{u}\cdot\nabla$ through \Eqns{\ref{mass}}{\ref{induction}} giving 
\begin{equation}
 \dyad{P} \colon \Grad\vec{u} =\frac{\ppar}{\rho}\frac{d\rho}{dt}-\Delta B \frac{dB}{dt}=\rho \frac{dU}{dt} \label{work}
\end{equation}
(noting that under most circumstances in this work, we will keep all thermodynamic quantities in units of energy density per unit mass).  In the guiding centre model, relationships between the moments of the distribution function are obtained from a kinetic equation \cite{Grad1966}. These relationships depend on the form of the distribution function rather than the macroscopic variables alone, which is inconvenient for a fluid description. However, a thermodynamic equation of state can be reconciled  \cite{Iacono1990,DeBlank1990} with the kinetic model and give the same results for certain simple choices of the functional form of the macroscopic variables. For example, an isotropic Maxwellian distribution, $p=\rho T(\psi)$, gives identical results in both models. In this study, our GCP compatible choice is to assume a two-temperature Maxwellian distribution  characterised by temperatures in the form
\begin{eqnarray}
\ppar(\rho,B,\psi)=\rho \Tpar(\psi) \label{tpar}\\
\pperp(\rho,B,\psi)=\rho \Tperp(B,\psi)  \label{tperp}
\end{eqnarray}
The functional dependence of $\Tperp$ on B is a result of \Equation{work}.
\section{Grad-Shafranov equation with toroidal flow and anisotropy}
Many examples of the Grad-Shafranov equation exist for static \cite{Grad1958}, anisotropic \cite{Grad1967,Zwingmann2001,Pustovitov2010}, and flowing \cite{Lovelace1986,Zehrfeld1972,Semenzato1984,Hameiri1983} equilibria. An equation incoporating both flow and anisotropy  \cite{Iacono1990,DeBlank1990} is used in this study and is outlined in this section.  Various reductions of this general case to other known forms are given in the appendix.

The basic scalar equations for force balance are obtained from components of \Equation{force}. The toroidal component of force balance yields a new flux function $F(\psi)$ for the toroidal magnetic field
\begin{eqnarray}
F(\psi)=(1-\Delta) RB_\phi \label{toroidalForce}
\end{eqnarray}
the parallel component of force balance gives a Bernoulli equation and new flux function $H(\psi)$
\begin{eqnarray}
H(\psi)=W(\rho,B,\psi)-\frac{1}{2}R^2\Omega(\psi)^2 \label{Bernoulli}\\
W(\rho,B,\psi)\equiv U+\frac{\ppar}{\rho} \label{enthalpy}
\end{eqnarray}
written in terms of a Legendre transform from energy $U$ to enthalpy $W$ in \Equation{enthalpy}.
Finally, we recover a modified Grad-Shafranov equation from the $\psi$ component of force balance
\begin{eqnarray}
\Div{\left [ { (1-\Delta) \left ( { \frac{\Grad\psi}{R^2}} \right ) } \right ]} =\nonumber \\
-\rho\Tpar'(\psi)- \rho H'(\psi) \nonumber\\
+\rho \left(\frac{\partial W}{\partial \psi}\right )_{B,\rho} - \frac{F(\psi)F'(\psi) }{R^2(1-\Delta)} + \rho R^2  \Omega(\psi)\Omega'(\psi) \label{GS}
\end{eqnarray}
subject to the integrability conditions constrained by energy conservation (\Equation{work})
\begin{eqnarray}
\left(\frac{\partial W}{\partial \rho} \right )_{B,\psi}=\frac{1}{\rho}\left(\frac{\partial \ppar}{\partial\rho}\right )_{B,\psi}\label{dwdrho} \\
\left(\frac{\partial W}{\partial B} \right )_{\rho,\psi}=\frac{1}{\rho}\left(\frac{\partial \ppar}{\partial B}\right )_{\rho,\psi}-\Delta(\rho,B,\psi) \frac{B}{\rho} \label{dwdb} 
\end{eqnarray}
Substituting the two-temperature Maxwellian GCP plasma expressions (\Eqns{\ref{tpar}}{\ref{tperp}}) into the integrability conditions (\Eqns{\ref{dwdrho}}{\ref{dwdb}}) gives a  choice for enthalpy $W_{\text{GCP}}(\rho,B,\psi)$ which is consistent with the GCP theory
\begin{eqnarray}
W_{\text{GCP}}(\rho,B,\psi)=\Tpar(\psi)\ln \left(\frac{\rho}{\rho_0} \frac{ \Tpar(\psi)}{ \Tperp(B,\psi)}\right ) + H_{\text{gauge}}(\psi) \label{Wgcp}\\
\Tperp(B,\psi)=\frac{B \Tpar(\psi)}{\left | B-\Tpar(\psi)\Theta(\psi)\right |} \label{ThetaGCP}
\end{eqnarray}
for arbitrary choice of $\rho_0$ and $ H_{\text{gauge}}$. The gauge transformation is possible because of the freedom to choose $\left( \partial W/\partial \psi \right )_{B,\rho}$ and is physically related to setting an arbitrary reference energy in the Bernoulli equation (\Equation{Bernoulli}). The integrability equations (\Eqns{\ref{dwdrho}}{\ref{dwdb}}) imply a flux function $\Theta$ which is a measure of anisotropy. With the above assumption about parallel heat transport, the system of equations is closed, and \Equation{GS} is now a second-order partial differential equation fully specified by the five flux functions:
\begin{equation}
\left \{\Tpar(\psi),H(\psi),\Omega(\psi),F(\psi),\Theta(\psi)\right \} \label{fluxfunctions}
\end{equation}
with definitions coming from \Equation{tpar}, \Equation{Bernoulli}, \Equation{flowfrozen}, \Equation{toroidalForce} and \Equation{ThetaGCP}.
The tokamak equilibrium problem has been reduced to solving \Equation{GS} by fitting the flux functions (\Equation{fluxfunctions}) to experimental data for appropriate boundary conditions.
\section{EFIT TENSOR}
\subsection{EFIT}
Codes based on the original EFIT reconstruct the toroidal current profile from experimental measurements or given values, assuming a radial MHD force balance parameterisation for the current \cite{Lao1985}. The Grad-Shafranov equation for isotropic and static cases is given by
\begin{equation}
\Delta^*\psi\equiv R^2 \Div \left ( \frac{\Grad\psi}{R^2}\right )=-R^2p'(\psi)-FF'(\psi)\label{GSStatic}
\end{equation}
which is the limiting case of the more general system (\Equation{GS}) for $\Omega(\psi)=0$~and~$\Theta(\psi)=0$ (although toroidal flow can be approximately incorporated into `pressure' in some codes, see appendix). The covariant representation for the field (\Equation{covB}) and Amp\`{e}re's law (\Equation{ampere}) give the identity
\begin{equation}
\Delta^*\psi=-RJ_\phi \label{jphiIdent}
\end{equation}
which is true for any 2-D axisymmetric equilibrium. This implies a parameterisation for the toroidal current
\begin{equation}
J_\phi(R,\psi)=Rp'(\psi)+FF'(\psi)/R \label{jphiStatic}
\end{equation}
when isotropic and static force balance is assumed. The two flux functions are decomposed into a linear combination of basis functions with constant coefficients
\begin{eqnarray}
p'(\psi)\approx\Sum{i}{1}{n_p} p_i b_{pi}(\psi) \\
FF'(\psi)\approx\Sum{i}{1}{n_f}  f_i b_{fi}(\psi) 
\end{eqnarray}
where the $b(\psi)$ represent the basis functions which are \emph{a priori} assumed. The EFIT++ code currently supports polynomial, tension spline or Chebyshev polynomial representations. \Equation{jphiIdent} is solved by alternately fitting $J_\phi$ to experimental or modelled constraints then performing a fast Buneman inversion \cite{Lackner1976} of
\begin{eqnarray}
\Delta^*(\psi^{(n+1)})=-RJ_\phi(R,\psi^{(n)}) \label{jphiIdentStatic}
\end{eqnarray}
at each $n$th iteration. The resulting $\psi(R,Z)$ solution is completely consistent with the input $J_\phi(R,Z)$, but a recalculation of the force balance criterion \Equation{jphiStatic} changes $J_\phi(R,Z)$ from the previous iteration. Thus, a self-consistent Picard iteration occurs between $\psi$ and $J_\phi(R,\psi)$ in \Equation{jphiIdentStatic} until the change in $\psi(R,Z)$ is below an arbitrary threshold. 
The free flux functions are fitted to the experimental and/or other indirectly calculated or assumed (`given') constraints with error by minimising
\begin{equation}
\chi^2=\Sum{i}{1}{N_M}\left(\frac{M_i-C_{iM}}{\sigma_{iM}} \right)^2+\Sum{i}{1}{N_G}\left(\frac{P_i-C_{iG}}{\sigma_{iG}} \right)^2
\end{equation}
where the $N$ are the number of data points, the $\sigma$ are the uncertainty, the $C$ are the calculated values and $M$ and $P$ are the measured and given values respectively. 
In addition to the plasma currents in a tokamak, other specified, induced or unknown currents  are present. When including these known, unknown and plasma currents, the calculated poloidal magnetic field at a given magnetic probe location is a superposition of current contributions through a linear combination of the current coefficients
\begin{eqnarray}
C^{(n+1)}_{iM}=\sum_{j=\text{known}} G(\vec{r}_i,\vec{r}_j)I_j \Newline
+\sum_{k=\text{free}} G(\vec{r}_i,\vec{r}_k)I^{(n+1)}_k \Newline 
+\int_{\text{plasma}}G(\vec{r}_i,\vec{r})J_\phi(R,\psi^{(n)})dRdZ \label{magnetics}
\end{eqnarray}
where $G$ is the response function for the probe. Other constraints are also expressed as a linear combination of the flux functions. Thus, the free coefficients and unknown currents constitute a linear least-squares problem expressible as
\begin{equation}
\chi^2=\Norm{A\vec{u}-\vec{k}}
\end{equation}
where $\vec{u}$ contains the unknown coefficients and currents, and $\vec{k}$ contains the constraints. For polynomial basis functions, the problem is solved in EFIT using singular value decomposition (SVD).
\subsection{EFIT TENSOR system of equations}
Here, we explicitly present the EFIT TENSOR system of equations in S.I. units
\begin{eqnarray}
\Delta^*\psi=-\mu_0RJ_\phi \\
J_\phi=R \frac{k}{m}\rho \Tpar'(\psi) + \frac{1}{\mu_0R(1-\Delta)}FF'(\psi) +\rho R^3\Omega\Omega'(\psi) \Newline
+R\rho H'(\psi)-R\left[\rho\left(\frac{\partial W}{\partial \psi}\right)_{\rho,B}+\frac{1}{\mu_0}\Div\left( \frac{\Delta}{R^2}\Grad\psi\right) \right] \label{jphiEfit}\\
\rho=\frac{\Tperp}{\Tpar(\psi)}\rho_0 \Exp{\frac{mH(\psi)}{k\Tpar(\psi)}}\Exp{\frac{m R^2\Omega^2(\psi)}{2k\Tpar(\psi)}} \label{rho}\\
W=\frac{k}{m}\Tpar(\psi)\Ln{\frac{\rho}{\rho_0}\frac{\Tpar(\psi)}{\Tperp}} \\
\Tperp=\frac{B\Tpar(\psi)}{\Abs{B-\Tpar(\psi)\Theta(\psi)}}\\
\Delta\equiv\mu_0 \frac{\ppar-\pperp}{B^2} \label{delta}
\end{eqnarray}
We have re-arranged \Equation{GS} into an expression in $J_\phi$ taking advantage of the very general \Equation{jphiIdent}.  We have also re-arranged the Bernoulli relation (\Equation{Bernoulli}) into an explicit form for the mass density $\rho$. For a given $\rho$, the current is almost a linear combination of the flux functions $\Tpar',\Omega\Omega',FF'$ and a non-linear function of $\Theta$.
This system of equations is subject to the following identities for the five free flux functions
\begin{eqnarray}
\Tpar(\psi)=\frac{\ppar}{\rho}\frac{m}{k} \label{identT} \\
\Omega(\psi)=\frac{v_\phi}{R} \label{identO}\\
F(\psi)=RB_\phi\left(1-\mu_0 \frac{\ppar-\pperp}{B^2}\right) \label{identF}\\
H(\psi)=\frac{\ppar}{\rho}\Ln{\frac{\rho}{\rho_0}\frac{\ppar}{\pperp}}-\frac{1}{2}v_\phi^2 \label{identH}\\
\Theta(\psi)=\frac{k}{m}\rho B\left(\frac{1}{\ppar}-\frac{1}{\pperp}\right) \label{identTh}
\end{eqnarray}
These identities are straightforward functions of the kinetic moments $\rho,v_\phi,\pperp,\ppar$ and field at any $(R,Z)$ location. The MHD particle mass $m$ and Boltzmann constant $k$ are only included to give $\Tpar$ dimensions of temperature, but  a more useful quantity is obtained when $k=1,m=1$ and $\Tpar$ has the interpretation of the ratio of parallel pressure $\ppar$ to mass density $\rho$. 
\subsection{Numerical scheme}
The EFIT TENSOR code is a significant alteration to EFIT with a completely different set of physical assumptions, equations and free functions. However, many of the methods and constraints were adaptable to the more general case, and here we describe those adaptations.

The most important difference characteristic of the more general system is that $J_\phi$ (\Equation{jphiEfit}) cannot in general be expressed as $J_\phi=J_\phi(R,\psi)$, nor can it be completely expressed as a linear combination of free flux functions. The problematic contributions are due to $\Delta,\rho$ and $\left(\partial W/\partial \psi\right)_{\rho,B}$. We parametrize the linear current terms and also $\Theta(\psi)$ in terms of the basis functions $b(\psi)$
\begin{eqnarray}
\Tpar'(\psi)\approx\Sum{i}{1}{n_t} t_i b_{ti}(\psi) \\
FF'(\psi)\approx\Sum{i}{1}{n_f}  f_i b_{fi}(\psi) \\
\Omega\Omega'(\psi)\approx\Sum{i}{1}{n_\omega}  \omega_i b_{\omega i}(\psi) \\
H'(\psi)\approx\Sum{i}{1}{n_h} h_i b_{hi}(\psi) \\
\Theta(\psi)\approx\Sum{i}{1}{n_\theta} \theta_i b_{\theta i}(\psi)
\end{eqnarray}
which are included in the iteration scheme
\begin{eqnarray}
J_\phi\Iter{n+1}=R \frac{k}{m}\rho\Iter{n}\Deby{\psi}\Tpar \Iter{n+1}  \Newline
+ \frac{1}{\mu_0R(1-\Delta\Iter{n})}\Deby{\psi}FF\Iter{n+1}   \Newline
+\rho\Iter{n} R^3\Deby{\psi}\Omega\Omega\Iter{n+1} \Newline
+R\rho\Iter{n}\Deby{\psi}H\Iter{n+1}-R\lambda\Iter{n+1} \Lambda (R,Z)\Iter{n} \label{jphiNonlinear}
\end{eqnarray}
where we have introduced a non-linear plasma current $\Lambda (R,Z)$ and weighting $\lambda$ corresponding to the bracketed term in \Equation{jphiEfit}. 
\begin{equation}
\Lambda (R,Z)\equiv\left[\rho\left(\frac{\partial W}{\partial \psi}\right)_{\rho,B}+\frac{1}{\mu_0}\Div\left( \frac{\Delta}{R^2}\Grad\psi\right) \right]
\end{equation}
This contribution is treated as a new basis function computed with bi-cubic spline derivatives with an associated coefficient $\lambda$ which is equal to unity for a converged solution. The reason in incorporating this non-linear term in this way instead of defining a fixed current was to allow greater flexibility for intermediate iterations at the cost of an additional degree of freedom. Forcing $\lambda=1$ would be equivalent to specifying a known current based on the previous iteration.  Contributions to the current from $\rho$ and $\Delta$ in \Equation{jphiNonlinear} are calculated from the previous iteration using \Eqns{\ref{rho}}{\ref{delta}}.

It is clear from \Equation{jphiNonlinear} that, in addition to the iteration of $\psi$ in \Equation{jphiStatic}, the current which satisfies force balance must also Picard iterate $\rho$, $B$ and $\Delta$ contributions. Of particular concern is the exponential dependance of the density $\rho$ on the flux functions $H,\Tpar$, and $\Omega$, which could conceivably produce large variation in the inferred flux functions with each iteration if not suitably constrained.

\Equation{jphiNonlinear} is a linear expression in terms of basis functions which can be inserted into standard EFIT current constraints such as \Equation{magnetics}. In addition, the flux functions may be independently constrained using the identities in \EqnsMany{\ref{identT}}{\ref{identTh}}. These identity constraints may be \emph{weak} or \emph{strong}; weak constraints depend on  the kinetic moments $\rho,v_\phi,\ppar$, or $\pperp$ and the intermediate calculation for $B$ whereas strong constraints also provide $B$. An additional constraint should be included which specifies that $\lambda=1$. The weighting on this condition may also be specified to aid with convergence.

An important feature of EFIT is the ability to infer the flux functions $p'(\psi)$ and $FF'(\psi)$ from the current profile. This inversion is possible due to the different $R$ dependence on the right-hand side of \Equation{jphiStatic}. However, this is not possible with the more complicated expression \Equation{jphiNonlinear}. The degeneracy of $\Tpar'(\psi)$ and $H'(\psi)$, with each term weighted equally in $R$ and $\rho$ means that the columns of the linear inversion are linearly dependant and impossible to distinguish from current measurements alone. Furthermore, the anisotropy flux function $\Theta(\psi)$ has no linear contribution to the current at all, but instead influences the current through $\rho$ and $\Delta$.  It is clear that EFIT TENSOR is a `forward code' when invoking flow and anisotropy physics with additional information required than the usual current constraints to resolve the degeneracy. This is the price paid for the fast linear inversion of an EFIT algorithm. A non-linear least-squares fit that includes density would break this degeneracy for significant computational cost and complexity. Given that tokamak equilibria need not be computed very often under many circumstances, this may be interesting to pursue in future work.

To date, EFIT++ supports a number of constraints,  and those that have been adapted to EFIT TENSOR are listed here for reference:  vacuum toroidal field, flux loops, magnetic probes, plasma current, poloidal field coils, safety factor on axis $q_0$, static and rotational pressure approximations, B components, diamagnetic flux, boundary, $\psi(R,Z)$ and Motional Stark Effect (MSE). Constraints available in other versions of EFIT for the magnetic field or current profile may be implemented in the same way on EFIT TENSOR with little modification. Most other kinetic constraints on temperature, density and rotation such as that provided by Thompson scattering and charge exchange recombination spectroscopy are currently absent from EFIT TENSOR, but can be included at present through the identity constraints on the five flux functions such as \Eqns{\ref{identT}}{\ref{identH}}.
\subsection{Dirichlet fixed-boundary mode}
The magnetostatic relation \Equation{jphiIdent} that is solved by EFIT at each iteration, constitutes a second-order inhomogenous partial differential equation on a finite discrete rectilinear domain with homogenous boundary conditions specified at infinity. For a given current distribution on that domain, the solution is unique. EFIT is a so-called `free boundary' code which iterates the current distribution on this domain to give the best fit to the (usually measured) constraints and external currents. The external currents are critical to the shape of the plasma boundary on this domain.

When considering analytical solutions, such as the \Solovev solution \cite{Solov'ev1975}, or other fixed-boundary problems, the external currents which give rise to the boundary are unknown. In particular, for the \Solovev case, the currents extend to infinity and cannot be included directly on a finite grid.  For the purposes of benchmarking against an analytical solution, a fixed-boundary mode was added to EFIT TENSOR. This was done using a `capacitance matrix' method (see for example \cite{Proskurowski1976}), which we will briefly describe here.

The `capacitance matrix' method is a way of calculating what surface currents are required on an arbitrary irregular boundary to satisfy a set of boundary conditions for $\psi$ on that boundary. Suppose we wish to solve \Equation{jphiIdent} on a subdomain $\omega$ bounded by $d\omega$ with inhomogenous Dirichlet boundary conditions $\psi(s)=f(s), J_\phi(s)=0, s\in d\omega$ for some arbitrary $f$. If we only care about the solution on the subdomain $\omega$, we may solve the problem using the Green's function from the infinite domain $\Omega$ using a sequence of steps:

First, solve the inhomogenous equation for $\psi_1$ on  $\Omega$ with boundary conditions at infinity
\begin{eqnarray}
\Delta^*\psi_1=-RJ_\phi
\end{eqnarray}
and obtain a new function $\psi_b$ by measuring $\psi_1$ on the boundary of the subdomain $d \omega$
\begin{eqnarray}
\psi_b(s)\equiv\psi_1(s), s\in d\omega
\end{eqnarray}
Next, define a different homogenous problem for $\psi_2$
\begin{eqnarray}
\Delta^*\psi_2=0
\end{eqnarray}
subject to the inhomogenous boundary condition using the function $\psi_b$ measured before $\psi_2(s)=f(s)-\psi_b(s)$. The function $\psi_2(s)$ may be expressed completely in terms of unknown surface currents $\hat{J_\phi}(s)$ and the Green's function $G(s',s)$ on $\Omega$
\begin{eqnarray}
\psi_2(s)=\int_{d\omega}  \hat{J_\phi}(s') G(s,s') ds' \nonumber \\
=f(s)-\psi_b(s)
\end{eqnarray}
For the discrete problem, the solution for the vector $\hat{J}_{\phi,i}$ may be expressed as the inversion of the response matrix $G$ containing only the points on the boundary
\begin{eqnarray}
G^{-1}_{ij}(f_j-\psi_{b,j})=\hat{J}_{\phi,i}
\end{eqnarray}
then the sum  $\psi=\psi_1+\psi_2$ uniquely satisfies the inhomogenous Dirichlet boundary conditions $\psi(s)=f(s), J_\phi(s)=0, s\in d\omega$ on $\omega$ and the fixed boundary problem is solved.

\begin{figure}
\begin{center}
\includegraphics{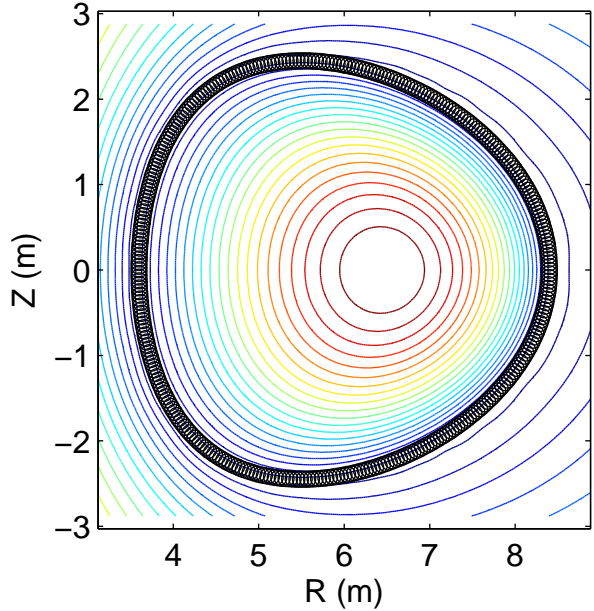}
\caption{A \Solovev solution as reconstructed with EFIT TENSOR using a fixed-boundary mode.}
\label{dirichletPic}
\end{center}
\end{figure}

In the current implementation of  EFIT TENSOR, the user specifies which grid points constitute the boundary and what the value of $\psi$ is on each grid point. Since the continuous boundary will not, in general, cross the grid points, the values at the grid points may vary accordingly. The last closed flux surface search is also disabled and specified as the supplied boundary. \FIGURE{dirichletPic} is an example fixed-boundary solution using this method. The $\psi$ contours inside the limiter region correspond to the \Solovev solution, and the contours outside the limiter are a by-product of the method and of no interest.
\section{Tests}
\subsection{Extension of \Solovev solution}
The \Solovev solution is an analytical solution to the Grad-Shafranov equation when $p(\psi)$ and $F^2(\psi)$ in \Equation{GSStatic}  are linear functions of $\psi$. Adopting a useful form used by \cite{Goedbloed2010}, the solution can be expressed as
\begin{eqnarray}
\PsiNorm=\SBrackets{x-\frac{1}{2}\epsilon\Brackets{1-x^2}}^2+ \nonumber \\
\Brackets{1-\frac{1}{4}\epsilon^2}\SBrackets{1+\epsilon\tau x\Brackets{2+\epsilon x}}\Brackets{\frac{y}{\sigma}}^2 \\
\psi=\Brackets{\frac{a^2B_0}{\alpha}}\PsiNorm \\
R=ax+R_0 \\
Z=ay \\
p_{S}(\PsiNorm)=p'[1-\PsiNorm]\\
F^2_{S}(\PsiNorm)=F'^2[1-\PsiNorm] +R_0^2B_0^2
\end{eqnarray}
where $\{\epsilon,\sigma,\tau\}$ control the shape of the $\psi$ solution and are known as the inverse aspect ratio, ellipticity and triangularity respectively. The field scaling is controlled by $\alpha$ and $B_0$. The linear profiles $\{p_S,F^2_S\}$ correspond to the static and isotropic flux functions and their gradients $\{p',F'^2\}$ depend on the above parameters. To make the simplest possible extension to flow and anisotropy, we re-write the Grad-Shafranov equation as
\begin{eqnarray}
\pperp=\pperp(R,B,\psi) \\
\Div{\left [ { \left ( { \frac{\Grad\psi}{R^2}} \right ) } \right ]} -\frac{F^2}{(1-\Delta)^2}\nabla_\psi\Ln{1-\Delta}= \nonumber \label{gradGS}\\ 
-\frac{1}{(1-\Delta) }\left (\frac{\partial \pperp}{\partial \psi} \right )_{B,R}  - \frac{F(\psi)F'(\psi) }{R^2(1-\Delta)^2}\\
\left (\frac{\partial \pperp}{\partial R} \right )_{\psi,B} = \rho R \Omega(\psi)^2 \label{flowCondition} \\
\left (\frac{\partial \pperp}{\partial B} \right )_{\psi,R} = -\Delta B \label{aniCondition}
\end{eqnarray}
If we wish to maintain the same $\psi$ geometry for our flow and anisotropy solution as for the \Solovev solution, then the partial derivatives of $\pperp$ and $F$ with respect to $\psi$ in \Equation{gradGS} must be the same, whilst also satisfying the additional conditions (\Eqns{\ref{flowCondition}}{\ref{aniCondition}}). By inspection, this is achieved with the profiles
\begin{eqnarray}
\pperp(R,B,\psi)=\frac{1}{2}\rho_0\Omega_0^2R^2-\frac{\Delta_0}{2}B^2+\sigma_0p_{S}(\psi) \label{solPress}\\
\ppar(R,B,\psi)=\frac{1}{2}\rho_0\Omega_0^2R^2+\frac{\Delta_0}{2}B^2+\sigma_0p_{S}(\psi) \\
F^2(\psi)=\sigma_0^2F^2_{S}(\psi)
\end{eqnarray}
where $\Omega_0,\sigma_0\equiv1-\Delta_0,\rho_0$ are constants with respect to $R,B,\psi$. On substitution of these expressions into \Equation{gradGS}, the logarithmic second term disappears and we re-obtain the ordinary Grad-Shafranov equation and the usual \Solovev solution. It is interesting to note that this implies that flow and/or anisotropy shear are required for non-\Solovev solutions with a linear pressure.  

There are advantages and disadvantages to this solution; although it exhibits the important de-coupling of magnetic surfaces and pressure surfaces, it does not do so with any respect for transport physics or thermodynamic assumptions. More specifically, one cannot write the pressures as $\ppar(\rho,B,\psi)=\rho \Tpar(\psi)$~and~
$\pperp(\rho,B,\psi)=\rho \Tperp(B,\psi)$ as assumed in EFIT TENSOR as discussed earlier. For example, dividing \Equation{solPress} by density $\rho_0$ will give $\Tperp=\Tperp(R,B,\psi)$, instead of the required  $\Tperp=\Tperp(B,\psi)$. This is interesting because it implies that the inclusion of flow and anisotropy in equilibrium introduces energy transport into the force balance inference problem. A related effect has been noted previously \cite{Iacono1990} when considering the direction of density shift for different enthalpy assumptions. It is, therefore, to be expected that when using this analytical solution as a constraint, the physical assumptions of the parallel heat transport model will prevent the same solution being obtained.  
\subsection{Comparison to analytical solution}
\subsubsection{Force balance convergence}
\begin{figure}[h]
\begin{center}
\includegraphics{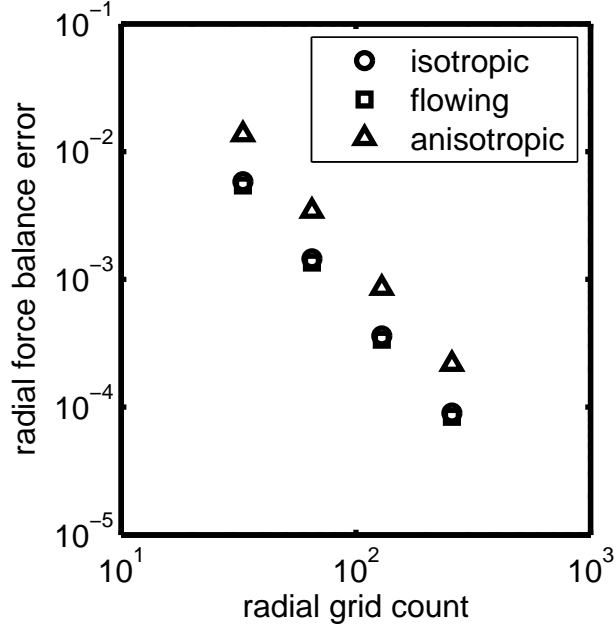}
\caption{ Comparison between plasma and field force balance at the geometric axis for extended \Solovev solutions. The parameters used for these cases are shown in Table~\ref{tableParams}.}
\label{convergence}
\end{center}
\end{figure}

A series of force balance numerical benchmarks were carried out on EFIT TENSOR by constraining to analytical \Solovev solutions and testing force balance of the resulting solution. A 2-point finite-difference comparison between the left and right hand sides of the fluid force equation (\Equation{force}) was used to measure force balance error. The linear current profile of the \Solovev solution meant that this force balance test was found to be accurate to one part in $1\E{-14}$ when the analytical solution was tested with the same scripts. 
The parameters used for the analytical constraints are shown in Table~\ref{tableParams}. The ordinary tokamak parameters such as field strength, plasma current and geometry were chosen to resemble an ITER like scale, but with the kinetic properties of flow, mass density and anisotropy exaggerated to maximize the changes to the analytical pressure from the ordinary case whilst still converging for the same numerical options such as grid resolution and polynomial order.
\begin{table}[htdp]
\caption{Extended \Solovev solution parameters}
\begin{center}
\begin{tabular}{|c|c|c|c|}
\hline
parameter & isotropic & flowing & anisotropic \\
\hline
$\epsilon$ & $0.4$& $0.4$ & $0.4$  \\
$\sigma$ & $1$ & $1$ & $1$ \\
$\tau$ & $1$ & $1$ & $1$ \\
$R_0$ & $6$~m & $6$~m& $6$~m\\
$B_0$ & $5$~T& $5$~T& $5$~T \\
$\alpha$ & $-3$ & $-3$& $-3$\\
$\rho_0$ & $1\E{-7}$& $1\E{-7}$& $1\E{-7}$ \\
$\Omega_0$ & 0  & $7\E{5}\text{rad s}^{-1}$ & 0 \\
$\Delta_0$ & 0 & 0 & $4\E{-3}$\\
$I_p$ & $16$ MA & $16$ MA & $16$ MA \\
$q^*$ & $1.6$ & 1.6 & 1.6 \\
$\beta_p$ & $1.0$  &2.5& 0.95\\
$\beta_T$ & $0.07$ &0.16 &0.06 \\
\hline
\end{tabular}
\end{center}
\label{tableParams}
\end{table}%
The results of the force balance benchmark are presented in \FIGURE{convergence}. The figure shows that the code accuracy scales equally well for static, flowing and anisotropic cases, with near identical accuracy for flow and static cases. The anisotropic cases were approximately 3 times worse than the static and flow cases, which is likely due to the spline derivatives used for the non-linear current calculation of $\Div\left( \frac{\Delta}{R^2}\Grad\psi\right)$. Better numerical evaluations of this term are perhaps possible, but have not been pursued in this work.
\subsubsection{Analytical solution reconstruction}
In this section, we compare the reconstructed solutions to the analytical solutions for the three cases of static isotropic, flowing and anisotropic plasma.  All three analytical solutions were generated on a 257$\times$257 grid. Kinetic moments $\ppar, \pperp, v_\phi~\text{and}~\rho$  were taken from the generated analytical solutions and used as input constraints for EFIT TENSOR. For the ordinary static case, the only free function was $\Tpar'$ with $n_t=1$ the order of the polynomial. For the flowing case, $H'$ was required for convergence and was specified as $n_h=2$. For the anisotropic case, $\Theta'$ was required instead, also with order $n_\theta=2$. These were the minimal order polynomials found to converge total $\psi$ to better than on part in $1\E{-15}$, $1\E{-11}$ and $1\E{-7}$ for the ordinary, flowing and anisotropic cases respectively. 

\begin{figure*}
\begin{center}
\includegraphics{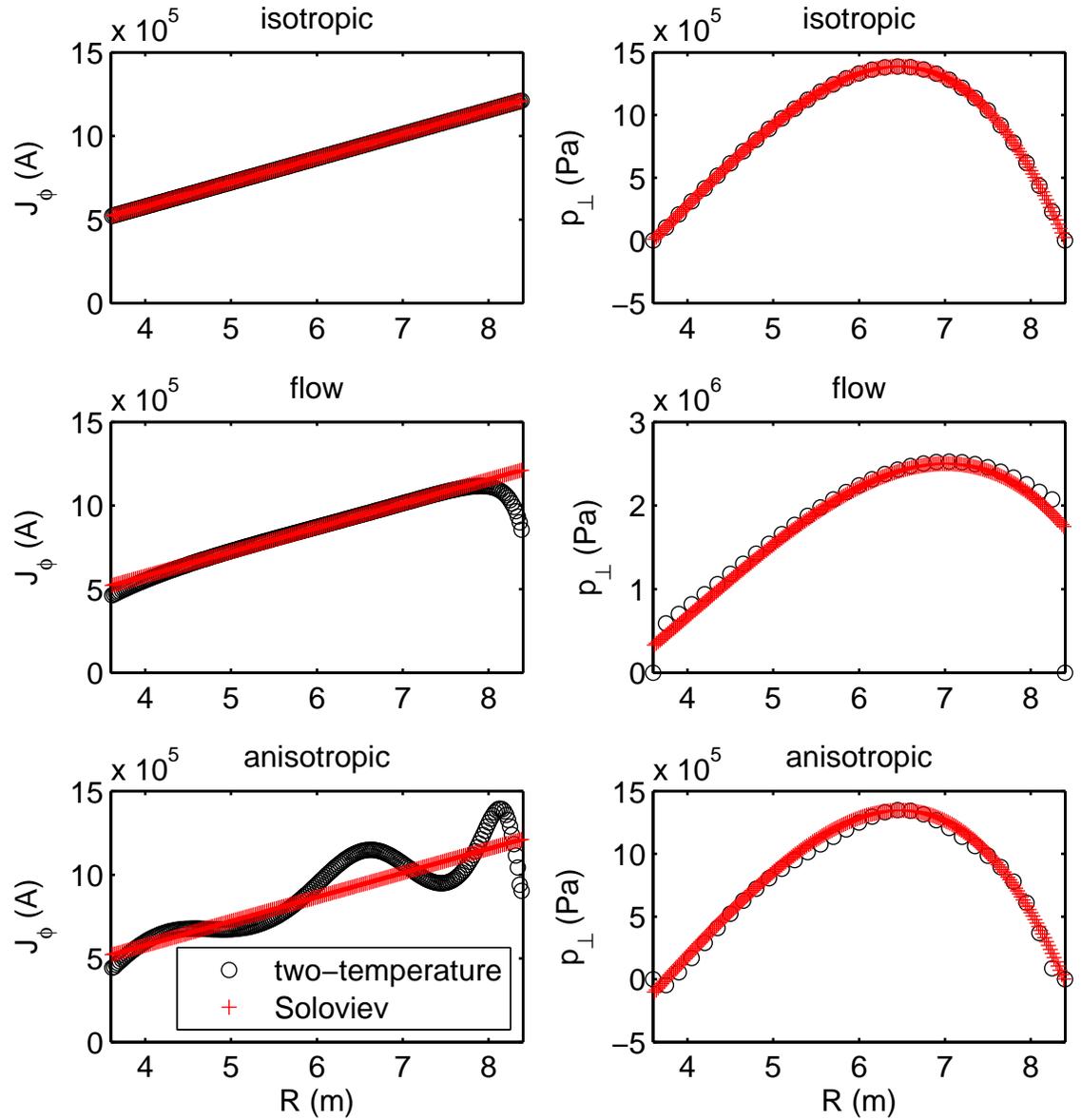}
\caption{Reconstruction of analytical radial profiles using EFIT TENSOR and constraining to pressure. The difference in parallel heat transport assumptions of the analytical solution and the two temperature GCP model has resulted in a different current profile for the same pressure profile.}
\label{constraintSOL}
\end{center}
\end{figure*}

A comparison of reconstructed radial current and pressure profiles is given in \FIGURE{constraintSOL} for each of the three cases. The only constraint required to obtain the static case was total current, whereas the flowing and anisotropic cases had identity constraints for $\Tpar, \Omega, H$ and $\Theta$ using \Eqns{ \ref{identT}}{\ref{identTh}}. No current or field profile data was used as an input in any case. The pressure profiles in \FIGURE{constraintSOL} show very good agreement; in the isotropic case, this is due to the uniqueness of equilibrium for a given current scaling assuming a linear pressure form. 

\begin{figure}
\begin{center}
\includegraphics{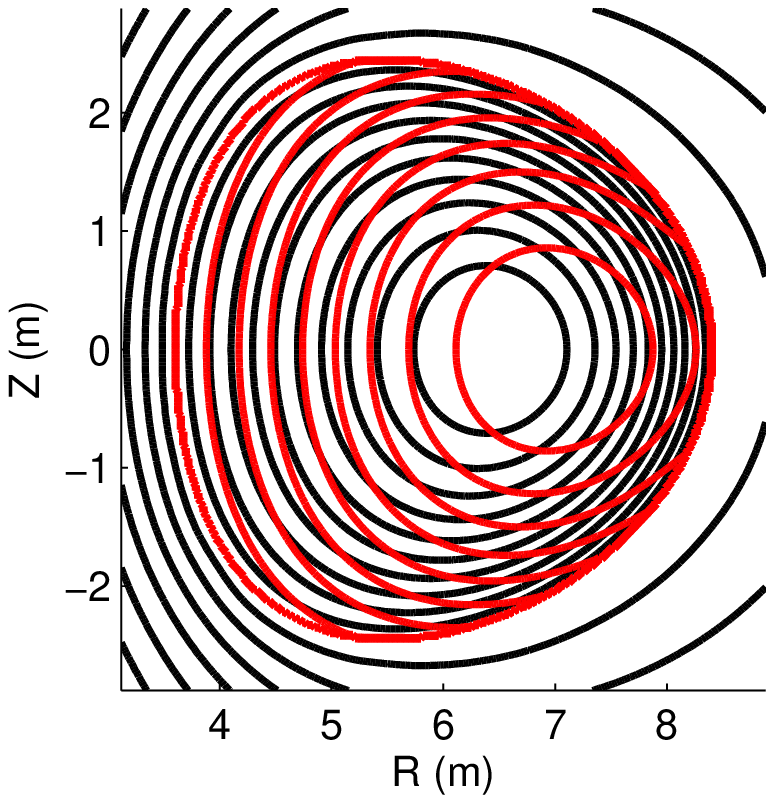}
\caption{The displacement of pressure contours from magnetic surfaces in the reconstructed flowing \Solovev benchmark. Pressure shown in red.} 
\label{flowDisplace}
\end{center}
\end{figure}

In the flow case, pressure was constrained, so it is unremarkable that the pressure profiles agree, however what is remarkable is that the pressure is displaced whilst the magnetic surfaces remain unchanged, as expected for large flows (corresponding to volume averaged sonic Mach number $0.9$ and \Alfven~Mach number $0.3$). \FIGURE{flowDisplace} is a plot of the EFIT TENSOR solution clearly showing the pressure displacement, whilst correctly satisfying force balance in \FIGURE{convergence}. As the flow increases with radius, the agreement between the reconstructed current and the analytical solution diverges as the heat flow assumption $T=T(\psi)$ becomes less relevant to the analytical case. 

The anisotropic case shows similar unremarkably good agreement for the constrained pressure, but a significant change is observed in the current profile. The disturbance is symmetric in poloidal flux $\psi$ and is a direct consequence of the unphysical $B^2$ dependence on $\Tpar$ in the analytical solution. Both the analytic solution and the EFIT TENSOR solution satisfy force balance but for different parallel heat flow assumptions in \Equation{GS}. This demonstrates that even with very good agreement for pressure profiles, the inferred magnetic topology can be radically different when considering anisotropy (even between anisotropy models), a result which qualitatively agrees with previous findings on $q$ profile in the presence of anisotropy \cite{Zwingmann2001,Hole2011}.  Conversely, it follows that any inference of pressure from a measured magnetic topology relies on assumptions about plasma transport and anisotropy.
\subsection{MAST TRANSP constraints}
\begin{figure}
\begin{center}
\includegraphics{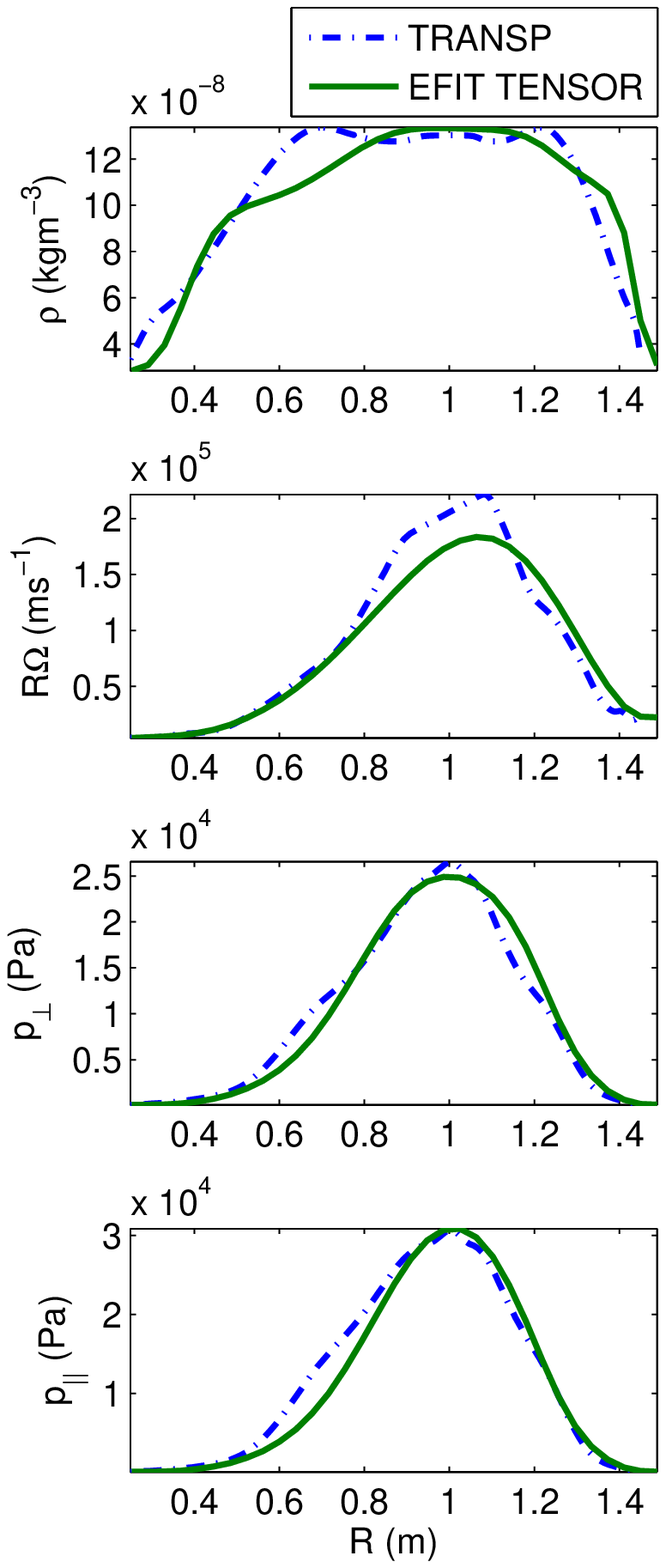}
\caption{Reconstructed radial profiles for MAST18696 at 290ms.}
\label{constraintTRANSP}
\end{center}
\end{figure}

\begin{figure*}
\begin{center}
\includegraphics{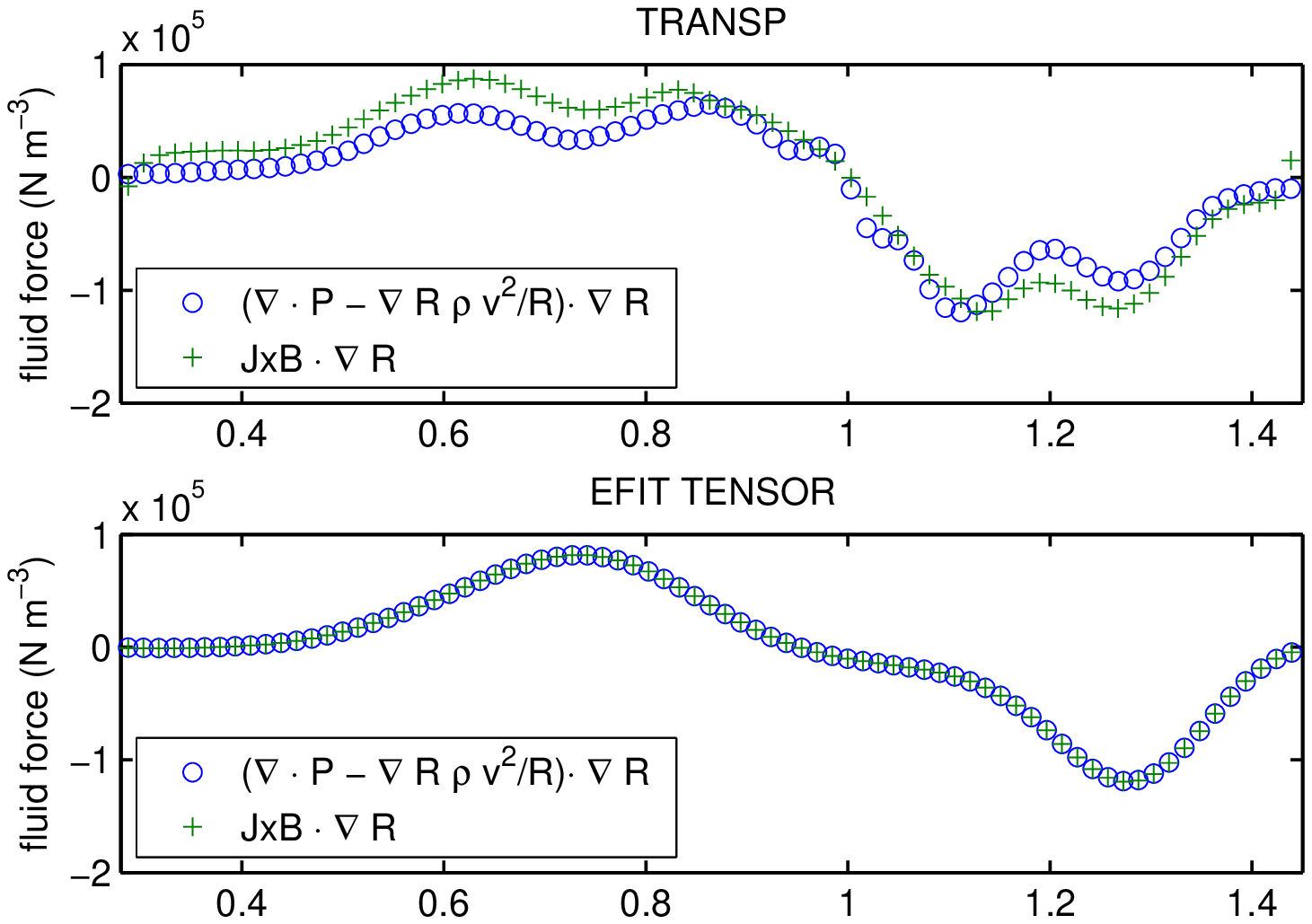}
\caption{Finite difference radial force balance check for profiles produced by TRANSP and EFIT TENSOR, with the inclusion of full order flow and anisotropy, for MAST discharge 18696 @ 290ms.}
\label{compare}
\end{center}
\end{figure*}

An equilibrium reconstruction was performed for MAST discharge 18696 at 290 ms using standard MAST EFIT++ magnetic constraints; magnetic probe, flux loops, field  and induced coil currents and total plasma current. Motional Stark Effect measurements were unavailable for this shot. In addition to the usual constraints, identity constraints on $\ppar,\pperp,v_\phi,\rho$ were provided from TRANSP \cite{Hawryluk1979} to constrain the five flux functions, thereby including full anisotropy and flow in the force balance. The TRANSP values were mapped to the mid-plane as functions of major radius. The resulting best fit is shown in \FIGURE{constraintTRANSP}. The largest discrepancy between TRANSP and EFIT TENSOR in this example is evident in the centrifugal force balance. Even though density and rotation were prescribed, an EFIT TENSOR solution with a lower rotation was inferred. This is likely related to the density being prescribed as a flux function in TRANSP, and the consequent density symmetry in radial mapping. The 2-D TRANSP magnetic field variables were extracted and used in a radial force balance benchmark similar to that used in previously in this paper (\FIGURE{compare}). The force balance test in \FIGURE{compare} shows that, for this example, TRANSP underestimates the plasma pressure contribution either side of the magnetic axis and that MHD equilibrium with flow and anisotropy is not satisfied away from the magnetic axis. A discrepancy is not surprising since TRANSP only uses a rotational pressure approximation and takes beam pressure to be the average of parallel and perpendicular components, but the magnitude of the error is interesting. The additional structure in the TRANSP force profile is due to the flat spots in the assumed TRANSP pressure profile in \FIGURE{constraintTRANSP}, which is not replicated with the second order polynomial used in this EFIT TENSOR example.

We have thus demonstrated an equilibrium reconstruction of a real MAST discharge including flow and anisotropy corrections to full order using measurements of radial profiles taken from TRANSP. We anticipate, in future work, to replace many of these constraints with experimental measurements of density, rotation and temperature.
\section{Conclusion}
A new force model for EFIT has been presented which includes arbitrary toroidal flow and anisotropy effects under the assumptions of ideal MHD for a single fluid guiding centre plasma. The parallel transport model assumed is that parallel temperature is a flux function. This model has been demonstrated and tested in the existing code EFIT++ and was found to produce identical results in the static and isotropic limits with a comparable computational cost. The model was also tested for force balance using a finite difference scheme and was benchmarked against an analytical solution. It was found that significant differences in the inferred current are possible for the same kinetic plasma constraints which result from the choice of parallel heat transport model. This implies that for sufficiently anisotropic plasma, the parallel transport model can be compared against measured current profiles, providing a novel measure of heat flow from equilibrium constraints.

In future work, we will investigate the effects of flow and anisotropy on MAST shots using EFIT TENSOR.  We anticipate that rotation and density constraints will be measured from experiment, and parallel and perpendicular pressures will be calculated using an NBI particle model such as NUBEAM \cite{Pankin2004} or LOCUST \cite{Akers2002}. It is anticipated that EFIT TENSOR will be merged back into EFIT++, replacing the isotropic Grad-Shafranov model.
\section{Appendix}
\subsection{Reduction of modified Grad-Shafronov equation to isotropic and/or static cases}
It is instructive to consider the reduction of the more general equilibrium problem to simpler cases, and the corresponding reduction in the number of free parameters. Indeed, treatments including poloidal flow have a free flux function in addition to the five presented in this work (but, it should be noted, for a different expression of $W(\rho,B,\psi)$ \cite{Iacono1990,DeBlank1990}).

Setting $\Delta\equiv 0$ in \Equation{GS} gives the isotropic force balance problem for toroidal flow
\begin{eqnarray}
\Div{\left [  \left ( { \frac{\Grad\psi}{R^2}} \right ) \right ]} =\nonumber \\
-\rho T'(\psi)- \rho H'(\psi) \nonumber\\
+\rho \left(\frac{\partial W}{\partial \psi}\right )_{\rho} - \frac{F(\psi)F'(\psi) }{R^2} + \rho R^2  \Omega(\psi)\Omega'(\psi) \\
H(\psi)=T(\psi)\ln \left(\frac{\rho}{\rho_0} \right )-\frac{1}{2}R^2\Omega(\psi)^2  \label{isoBernoulii} \\
W(\rho,\psi)=T(\psi)\ln \left(\frac{\rho}{\rho_0} \right )\\
\left \{T(\psi),H(\psi),\Omega(\psi),F(\psi) \right \}
\end{eqnarray}
which is the form found by many authors \cite{Lovelace1986,Zehrfeld1972,Semenzato1984,Hameiri1983} . A consequence of \Equation{isoBernoulii} is another widely known and useful form for this system
\begin{eqnarray}
\Div{\left [  \left ( { \frac{\Grad\psi}{R^2}} \right ) \right ]} = -\left (\frac{\partial p}{\partial \psi} \right )_R - \frac{F(\psi)F'(\psi) }{R^2}  \label{partialflowGS}\\
\left (\frac{\partial p}{\partial R} \right )_\psi = \rho R \Omega(\psi)^2 \\
p(R,\psi) \equiv \rho(R,\psi)T(\psi) \\
\left\{ p(R,\psi),\Omega(\psi),F(\psi) \right\}
\end{eqnarray}
If we further assume zero flow by setting $\Omega(\psi)\equiv 0$ in \Equation{GS} then, on re-examination of  \Equation{isoBernoulii},  we identify no more explicit radial dependence and a redundant separation of temperature and density. Thus, we arrive at the basic Grad-Shafranov equation
\begin{eqnarray}
\Div{\left [  \left ( { \frac{\Grad\psi}{R^2}} \right ) \right ]} = -p'(\psi) - \frac{F(\psi)F'(\psi) }{R^2}  \\
H(\psi)=T(\psi)\ln \left(\frac{\rho(\psi)}{\rho_0} \right ) \\
p(\psi)\equiv\rho(\psi)T(\psi)\\
\left \{p(\psi),F(\psi) \right \}
\end{eqnarray}
Conversely, the purely anisotropic system becomes
\begin{eqnarray}
\Div{\left [ { (1-\Delta) \left ( { \frac{\Grad\psi}{R^2}} \right ) } \right ]} =\nonumber \\
-\rho\Tpar'(\psi)- \rho H'(\psi) \nonumber\\
+\rho \left(\frac{\partial W}{\partial \psi}\right )_{B,\rho} - \frac{F(\psi)F'(\psi) }{R^2(1-\Delta)}\\
H(\psi)=\Tpar(\psi)\ln \left(\frac{\rho}{\rho_0} \frac{ \Tpar(\psi)}{ \Tperp(B,\psi)}\right ) \\
W(\rho,B,\psi)=\Tpar(\psi)\ln \left(\frac{\rho}{\rho_0} \frac{ \Tpar(\psi)}{ \Tperp(B,\psi)}\right ) \\
\Tperp(B,\psi)=\frac{B \Tpar(\psi)}{\left | B-\Tpar(\psi)\Theta(\psi)\right |} \\
\left \{\Tpar(\psi),H(\psi),F(\psi),\Theta(\psi)\right \} 
\end{eqnarray}
and analogous to the pure flow case, we perform the full $\psi$ derivative of the Bernoulli equation and using the integrability relations (\Eqns{\ref{dwdrho}}{\ref{dwdb}}) we obtain a similar set of equations
\begin{eqnarray}
\Div{\left [ { (1-\Delta) \left ( { \frac{\Grad\psi}{R^2}} \right ) } \right ]} =-\left (\frac{\partial \ppar}{\partial \psi} \right )_B  \nonumber \\
 - \frac{F(\psi)F'(\psi) }{R^2(1-\Delta)}\\
\left (\frac{\partial \ppar}{\partial B} \right )_\psi = \Delta B\\
\ppar(B,\psi) \equiv \rho(B,\psi)\Tpar(\psi)\\
\left\{ \ppar (B,\psi),\Theta(\psi),F(\psi) \right\}
\end{eqnarray}
which reduces immediately to the form given by Grad\cite{Grad1967}.
\subsection{Rotational pressure approximation}
Here we relate our set of free functions (\Equation{fluxfunctions}) to the `rotational pressure' approximation used in some existing codes (such as \cite{Lao2005} and \cite{Appel}). The definition of rotational pressure is through an expansion in major radius
\begin{eqnarray}
\left (\frac{\partial \ppar}{\partial \psi} \right )_R \approx P_{\text{axis}}'(\psi )+x P_{\text{rot}}'(\psi )\\
x\equiv\left(\frac{R^2}{R_0^2}-1\right)
\end{eqnarray}
where $R_0$ is some origin (ideally, the magnetic axis). Casting the isotropic pressure in terms of $x$ and Taylor expanding around $x=0$ gives the required definitions in S.I. units
\begin{eqnarray}
\ppar(x,B,\psi)= \rho_0 \frac{k}{m}\frac{B}{\Abs{B-\Theta(\psi)\Tpar(\psi)}}\nonumber\\ \times\Tpar(\psi )\Exp{ \frac{m H(\psi )}{k \Tpar(\psi )}}\nonumber\\
\times\Exp{\frac{m R_0^2 (x+1) \Omega (\psi )^2}{2 k \Tpar(\psi )}} \\
D_{\text{axis}}(\psi )\equiv \rho_0 \frac{B}{\Abs{B-\Theta(\psi)\Tpar(\psi)}} \Exp{ \frac{m H(\psi )}{k \Tpar(\psi )}}\nonumber\\
\times\Exp{\frac{m R_0^2 \Omega (\psi )^2}{2 k \Tpar(\psi )}}\\
P_{\text{axis}}(\psi )\equiv  \frac{k}{m}\Tpar(\psi )D_{\text{axis}}(\psi )\\
P_{\text{rot}}(\psi )\equiv\frac{1}{2} R_0^2 \Omega(\psi )^2 D_{\text{axis}}(\psi )
\end{eqnarray}
which is clearly only strictly appropriate for $\Theta(\psi)=0$, \IE~when the system is isotropic.
\section{Acknowledgments}
The first author would like to thank Luca Guazzotto (University of Rochester), Greg von Nessi (ANU), Rob Akers and Ken McClements (CCFE) for very useful discussions at various points in this research, and David Muir (CCFE) and David Pretty (ANU) for invaluable assistance with MAST data access. 
This work was funded by the Australian Research Council through Grant Nos. DP1093797 and FT0991899 and by the RCUK Energy Programme under Grant No. EP/I501045 and the European Communities under the contract of Association between EURATOM and CCFE. The views and opinions expressed herein do not necessarily reflect those of the European Commission.
\bibliographystyle{unsrt}
\bibliography{efit}

\end{document}